\input amstex
\magnification=\magstep1
\documentstyle{amsppt}

\NoBlackBoxes

%
%
\newcount\equanumber \equanumber=0
\def\eqlab #1%
    {%
    \LookUp Eq_#1 \using\equanumber%
    {\label}%
    }%
%
%
\newcount\eqauto    \eqauto=0
\def\eqnum{\global\advance\eqauto by 1
                  \eqlab{{\jobname.\the\eqauto}} }
%
%
\newcount\thmnumber \thmnumber=0
\def\thmlab #1%
    {%
    \LookUp Eq_#1 \using\thmnumber%
    {\label}%
    }%
%
%
\newcount\thmo    \thmo=0
\def\thmnum{\global\advance\thmo by 1
                  \thmlab{{\jobname.\the\thmo}} }
%
%
\newcount\refnumber \refnumber=0
\def\reflab #1%
    {%
    \LookUp Ref_#1 \using\refnumber%
    {\label}%
    }%
%
%
\def\LookUp #1 \using#2%
    {%
    \expandafter\ifx\csname#1\endcsname\relax
        \global\advance #2 by 1
        \expandafter\xdef\csname#1\endcsname{\number#2}%
    \fi
    \xdef\label{\csname#1\endcsname}%
    }%

\TagsOnRight
\font\script=rsfs12 at 10pt

\def\ie{{\it i.e.}}
\def\etc{{\it etc}}
\def\p{\partial}
\def\d{\text{d}}
\def\D{\Cal D}
\def\hor{\text{hor\,}}
\def\a{\alpha}

\def\H{\Cal H}
\def\g{\goth g}
\def\O{\Omega}
\def\Ham{\text{\script H}}
\def\eff{\text{eff}}

\def\P{\Cal P}
\def\Or{{\Cal O}}
\def\sgn{\text{sgn}}
\topmatter
\title 
Quantization of a particle in a background Yang-Mills field
\endtitle
\author Yihren Wu\endauthor
\affil Department of Mathematics, Hofstra University, Hempstead, NY 11550
\endaffil
\endtopmatter
\centerline{\bf Abstract}

Two classes of observables defined on the phase space
of a particle are quantized, and 
the effects of the Yang-Mills field are discussed in the context of
geometric quantization.


Short title: Particle in background Yang-Mills field.

\vfil\pagebreak
\subheading{I. Introduction}

Let $Q$ be a Riemannian manifold considered as
the configuration space of a particle, 
the purpose of  this paper is to discuss the 
quantization of the observables  on the phase space $T^*Q$
of this particle when it is moving under the influence
of a background Yang-Mills field, 
the Yang-Mills potential is 
a connection $\a$ on a principal bundle $N$ over $Q$.

The free $G-$action on $N$ can be lifted to a Hamiltonian $G-$action on
$T^*N$ with an equivariant moment map $J:T^*N\to\g^*$.
Let $\mu\in\g^*$, and denote by $\Or_\mu$ the coadjoint orbit 
through $\mu$.
Then $J^{-1}(\Cal O_\mu)/G$ has a canonical
symplectic structure given by the Marsden-Weinstein reduction
\cite{\reflab{MW}}.
This reduced phase space is the appropriate phase space of a particle in a 
background Yang-Mills field
$\a$ of charge $\mu$ \cite{\reflab{Weinstein}}.

We will denote by $\Cal Q(X)$ the quantization of the phase space
$X$, suppressing in our notation the
choices of polarizations and pre-quantization line bundles \etc,
via the standard procedure of geometric quantization \cite{\reflab{Sn},
\reflab{woodhouse}}.
Suppose we choose the vertical polarization
on $T^*N$ so that the quantization $\Cal Q(T^*N)$
of $T^*N$ gives $L^2(N)$. Moreover,
suppose the co-adjoint 
orbit $\Cal O_\mu$ is integral, so that
the quantization of this  coadjoint orbit  gives a
irreducible representation space $\H_\mu$ of $G$ \cite{\reflab{Ki}} 
\cite{\reflab{Ko}}. A theorem
of Guillemin-Sternberg \cite{\reflab{multi}} 
(see also \cite{\reflab{Robson}}) then suggests that the
quantization of $J^{-1}(\Cal O_\mu)/G$ is given by 
$\text{Hom}_G(\H_\mu, L^2(N))$,
the space of $G-$equivariant linear maps from $\H_\mu$ to $L^2(N)$. 
And this result holds independent of whether there is a Yang-Mills
field present in the backgroud.
Thus when some technical assumptions are made
so that the procedure of geometric quantization can be carried out 
smoothly,
the Yang-Mills field plays no role in
the quantization of the phase space 
$J^{-1}(\Or_\mu)/G$.

We will discuss the effect of the Yang-Mills field in 
quantizing observables that are lifted from functions on $T^*Q$.
We will show that the resulting quantum operators are expressed in 
terms of the covariant derivatives, which is defined by the
connection $\a$.
In particular, we will show that the quantum operators
for $f$ 
of the form $\tfrac{1}{2}||p||^2+V(q)$ are expressed in terms 
of the covariant Laplace operator and the Ricci curvature. 
This is obtained by a standard
Blattner-Kostant-Sternberg (BKS) pairing approach \cite{\reflab{BKS}}. 
Our results are in
agreement with those of Landsman \cite{\reflab{Landsman}} 
who arrived at the conclusion via deformation quantization.

Outline of this paper is as follows; In section 2 we give a detailed
exposition of our problem in order to standardize the notations
used throughout this paper. 
As a prelude to our result, we note that if the gauge group 
is abelian, we will recover the Dirac quantization of a 
charge particle in the presense of an electro-magnetic field. In section
3 we introduce local coordinates to facilitate our calculation, and state
some results concerning the Hamiltonian vector fields for our observables.
We follow closely the treatment of \cite{\reflab{Robson}} on the 
polarization chosen for the phase space in section 4. Our results when 
$f$ is polarization preserving are given in section 5, and the quantization
of $\tfrac{1}{2}||p||^2+V(q)$ using BKS pairing appears in section 6.

\subheading{II. Preliminary discussions}

Let $N$ be a principal $G-$bundle over $Q$ where $G$ is compact, with
Lie algebra $\g$,
and the group action is on the right. 
We define two functions 
$R_g:N\to N$ and $\hat n:G\to N$, where
$$R_g(n)=\hat n(g)=ng,\tag\eqlab{basic}$$
and we denote by $F_*$ the Jacobian of $F$.
A connection is a linear map $\a(n):T_nN\to\g$ for each $n\in N$ satisfying
\roster
\item"{\it i.}" $\text{Ad}_{g^{-1}}\a(n)=\a(ng)R_{g*}:T_nN\to\g$,
\item"{\it ii.}" $\a(n)\hat n_*=\text{Id}:\g\to\g$.
\endroster
The free $G-$action on $N$ can 
be lifted to a Hamiltonian $G-$action on $T^*N$ with moment map 
$J:T^*N\to\g^*$ given by $J(\xi,n)=\hat n^*\cdot \xi$.

Let $N^\#$ be the pullback
bundle over $T^*Q$:
Explicitly, 
$$N^\#=\{(p,n)\,|\,p\in T^*_qQ\text{ where }q=\pi(n)\}.$$
Define a diffeomorphism 
$$\align \chi:N^\#\times \g^* &\to T^*N,\tag{\eqlab{chi}}\\
(p,n,\mu) &\mapsto (\xi,n) \quad
\text{where}\quad\xi=\pi^*(p)+\langle\mu,\a\rangle
\in T^*_nN.
\endalign$$ 
This map in turn induces an $\a-$dependent projection $\pi_\a: T^*N\to T^*Q$,
and their corresponding symplectic forms are related by
$$\O_{T^*N}=\pi^*_\a\O_{T^*Q}+\d\langle\mu,\a\rangle.\tag{\eqlab{a}}$$
One shows that the moment map is simply the projection
$N^\#\times\g^*\to\g^*$, \ie, $\chi^{-1}(J^{-1}(\mu))=N^\#\times\{\mu\}$. 
Thus
for each $\mu$, the $G-$action on $N^\#$ induces a $G-$action on $J^{-1}(\mu)$.
This action is non-canonical ($R_g^*\,\O_{T^*N}\ne\O_{T^*N}$) in general: 
$$(\xi,n)\mapsto (\xi_g,ng)\quad\text{where}\quad\xi_g=R^*_{g^{-1}}\xi+
R^*_{g^{-1}}[(\text{Ad}^*_{g^{-1}}-\text{Id})\mu]\a(n).$$
However,
this action coincides with the canonical $G-$action when restricted to
the isotropy subgroup 
$$H=\{g\in G\,|\,\text{Ad}_g^*\mu=\mu\}$$ 
of $\mu$. (There
$\text{Ad}^*_{g^{-1}}\mu=\mu$ and $\xi_g=R^*_{g^{-1}}\xi$.) 
The relevant phase space becomes
$J^{-1}(\Cal O_\mu)/G=J^{-1}(\mu)/H=N^\#/H\times\{\mu\}$.

Given an observable on the phase space of the particle 
$f:T^*Q\to \bold R$,
by the projection $\pi_\a$, the pullback map, which we continue to denote
by $f$,
$$f=\pi_\a^*f:T^*N\simeq N^\#\times \g^*\longrightarrow\bold R,
\tag{\eqlab{fback}}$$
is invariant with respect to both the canonical $G-$action and the 
non-canonical one. In particular, $f$ is independent on the charge
variables $\mu\in\g^*$.

We assume  that $\Or_\mu$ is integral,
so that $\Cal Q(\Or_\mu)=\H_\mu$ is an irreducible representation space of
$G$ induced by $\rho_\mu:H\to U(1)$.
Choose an orthonormal basis $\{\phi_i\}$ for $\H_\mu$, where $\phi_i$ is a
holomorphic function on the K\"ahler manifold $G/H$ \cite{\reflab{Ki}}.
Then $\Psi\in\Cal Q(J^{-1}(\mu)/H)=\text{Hom}_G(\H_\mu,L^2(N))$ is determined
by $\Psi(\phi_i)=\Psi_i\in L^2(N)$. Using orthonormality of $\phi_i$ and
$G-$equivariance of $\Psi$,
we write
$$\Psi=\sum\Psi_i\phi_i:N\times G/H\to N\times_G G/H\to\bold C $$
which is uniquely determined by $\psi=\Psi(-,eH):N\to \bold C$ with
the condition
$\psi(nh)=\rho_\mu(h^{-1})\psi(n)$ for all $h\in H$.
So we see that the Yang-Mills potential plays no role in quantizing 
 the relevant phase space, it simply picks up the multiplicity of the
charge sector $\Cal O_\mu$ in $L^2(N)$
(cf. \cite{\reflab{multi}}, \cite{\reflab{Robson}}).

If we are to quantize an observable  that is a pullback of $f$ on
the phase space of the particle 
$T^*Q$, the connection $\a$ plays an important role.
As an illustration, suppose
the charge $\mu\in\g^*$ is $G-$invariant, then $H=G$. This will be the
case if for instance $G$ is abelian. Under this condition,
$J^{-1}(\mu)/H$ is diffeomorphic to $T^*Q$ via the projection $\pi_\a$,
and the canonical symplectic form on the reduction space pushes forward onto
$T^*Q$. So $J^{-1}(\mu)/H$ is symplectomorphic to $T^*Q$ if $T^*Q$ is
equipped with the ``effective'' symplectic form 
$\O_{\eff}=\O_{T^*Q}+\langle\mu,\O_\nabla\rangle$
where $\O_\nabla$ is a two-form on $Q$ which pulls back to the curvature
form $\d\a$ on $N$ \cite{\reflab{Kummer}}. 
It is with respect to this effective symplectic form
that the quantization procedure must be carried out. Since the adjustment
is a two form on $Q$, quantization of $T^*Q$ using the vertical polarization
still gives $L^2(Q)$. However, Hamiltonian vector fields $\Ham_{f_i}$
associated with
observables $f_i$ and the Poisson bracket are defined in terms of the 
effective form:
$$\O_{\eff}(\Ham_{f_i},-)=-\d f,\qquad\qquad
\{f_1,f_2\}=\O_{\eff}(\Ham_{f_1},\Ham_{f_2}).$$
The quantization of observables must preserve Poisson bracket
$$[\Cal Q(f_1),\Cal Q(f_2)]=i\hbar\Cal Q(\{f_1,f_2\}).
$$
When carry out the geometric quantization with respect to $\O_{\eff}$,
the result is the Dirac quantization with $\a$ as the vector potential
associated with a  electro-magnetic field 
(cf. \cite{\reflab{Sn}}).

\subheading{III. The Hamiltonian vector fields}

Let us first introduce local canonical coordinates $(\xi_i,n_i)$ on $T^*N$ so
that $\O_{T^*N}=\d\xi_i\d n_i$, similarly $(p_a,q_a)$ on $T^*Q$
with $\O_{T^*Q}=\d p_a\d q_a$, and let $\a=A_{si}\d n_i$ where 
$i=1\dots\dim N$, $a=1\dots\dim Q$,
$s=1\dots\dim G$, and repeated indices are summed. 
For each $n\in N$, let us denote the horizontal
lift $T_qQ\to T_nN$ by the matrix $M_{i\sigma}(n)$, $i=1\dots\dim N,\ 
\sigma=1
\dots\dim Q$. We have 
$$\frac{\p q_a}{\p n_i}M_{i b}=\delta_{ab}\tag{\eqlab{lift}}$$
and the covariant derivative is
the horizontal lift of $\tfrac{\p\ }{\p q_a}$:
$$\D_a=M_{ia}\frac{\p\ }{\p n_i}.\tag{\eqlab{covariant}}$$
In these coordinates, the canonical one-form and the symplectic two-form
of $T^*N$ can be calculated using (\eqlab{a})
$$\align
\chi^*\xi_i\d n_i&=\left(p_a\frac{\p q_a}{\p n_i}+\mu_sA_{si}\right)\d n_i\\
\chi^*\O_{T^*N}&=\frac{\p q_a}{\p n_i}\d p_a\d n_i+
\mu_s\frac{\p A_{si}}{\p n_j}\,\d n_j\d n_i
+A_{si}\,\d\mu_s\d n_i.\tag{\eqlab{localform}}
\endalign$$
Let $f:N^\#\times \g^*\to\bold R$ be a pullback function from $T^*Q$, and
let $\Ham_f$ be its Hamiltonian vector field
$$\Ham_f=B_a\frac{\p\ }{\p p_a}+C_i\frac{\p\ }{\p n_i}+
U_s\frac{\p\ }{\p\mu_s}\,.$$
Using $\chi^*\O_{T^*N}(\Ham_f,-)=-\d f$ we get
$$\align\frac{\p f}{\p q_a}\frac{\p q_a}{\p n_i}
&=-B_a\frac{\p q_a}{\p n_i}+\mu_s\left(\frac{\p A_{sj}}{\p n_i}-
\frac{\p A_{si}}{\p n_j}\right)C_j-A_{si}U_s,\\
\frac{\p f}{\p p_a}&=\frac{\p q_a}{\p n_i}C_i,\tag{\eqlab{Ci}}\\
\frac{\p f}{\p \mu_s}&=A_{si}C_i.
\endalign$$
Since $f$ is invariant with respect to the canonical $G-$action, $\Ham_f$ is
tangent to the subspace $J^{-1}(\mu)\simeq N^\#\times\{\mu\}$, thus $U_s=0$.
As remarked after (\eqlab{fback}), $f$ is independent of $\mu$, thus 
$A_{si}C_i=0$, which implies $\Ham_f$ is horizontal. 
Moreover, letting $B_a=-\frac{\p f}{\p q_a}+E_a$, we have
$$\Ham_f=\left[-\frac{\p f}{\p q_a}\frac{\p\ }{\p p_a}
+C_i\frac{\p\ }{\p n_i}\right] + E_a\frac{\p\ }{\p p_a}\,.
$$
The terms in the bracket is the horizontal lift of
the Hamiltonian vector field of $f$ with respect to the
usual symplectic form $\O_{T^*Q}$ on $T^*Q$. $E_a$ satisfies
$$E_a\frac{\p q_a}{\p n_i}=\mu_s\left(\frac{\p A_{sj}}{\p n_i}-
\frac{\p A_{si}}{\p n_j}\right)C_j.$$
So we summarize the properties of $\Ham_f$ needed for our purpose.
\proclaim{Proposition \thmlab{function}} If $f:T^*N\to \bold R$ is a 
pullback of a function  on $T^*Q$, then for all $\mu\in\g$
the Hamiltonian vector field
$\Ham_f$ is a vector field on the subspace $J^{-1}(\mu)$, and as  the
total space of a principal bundle over $T^*Q$ with connection $\a$, 
$\Ham_f$ is horizontal. If differs from the horizontal lift of the
standard Hamiltonian vector field on $T^*Q$ by a field in the vertical
direction. With respect to the local coordinates chosen, we have
explicitly:
$$\Ham_f=\left[-\frac{\p f}{\p q_a}\frac{\p\ }{\p p_a}
+M_{ia}\frac{\p f}{\p p_a}\frac{\p\ }{\p n_i}\right]
+\mu_sM_{ia}M_{jb}
\left(\frac{\p A_{sj}}{\p n_i}-\frac{\p A_{si}}{\p n_j}\right)
\frac{\p f}{\p p_b}\frac{\p\ }{\p p_a}\,.\tag{\eqlab{explicitham}}$$
Furthermore, we have
$$\langle \chi^*\xi_i\d n_i,\Ham_f\rangle=p_a\frac{\p f}{\p p_a}\,.
\tag{\eqlab{theta}}$$
\endproclaim

\subheading{IV. Polarization}

We first state some well known results concerning  the 
quantization of integral coadjoint 
orbits 
$\Cal O_\mu$. Let $\goth h$ be the Lie algebra of $H$,
we say that 
$\Or_\mu$ is integral if the map
$$v\in\goth h\to 2\pi i\langle v,\mu\rangle\tag\eqlab{integral}$$
is the derivative of a global character, \ie, there is a group homomorpism
$\rho_\mu:H\to U(1)$ such 
that $\rho_{\mu*}$ is the map given in (\eqlab{integral}). 
A version of the Borel-Weil theorem, 
due to Kirillov \cite{\reflab{Ki}} and Kostant \cite{\reflab{Ko}}
asserts that there is a one-to-one correspondence between the integral
orbits of $G$ and its unitary irreducible representations, and these 
representations can be construction by the method of geometric 
quantization applied to the coadjoint orbit $\Or_\mu$, which we will briefly
explain.

Let $\Cal L$ be the prequantization
line bundle $G\times_{\rho_\mu}\bold C$ over $\Cal O_\mu\simeq G/H$ with
connection induced by the map $\rho_\mu$.
It is known that $\Cal O_\mu$ is a
K\"ahler manifold with complex coordinates 
with respect to which the $G-$action is holomorphic. There is a standard
$G-$equivariant polarization quantizing with respect to the  
line bundle and this polarization gives $\H_\mu$ whose 
elements are holomorphic
functions on $\Cal O_\mu$. The polarization, known as the positive
K\"ahler polarization, is given by left translation of a set of
$v_k\in\g\otimes\bold C$, the complexification of $\g$, so that the
polarization is generated by 
$$V_k(\text{Ad}_g\mu)=g_*v_k\in T_{\text{Ad}_g\mu}\Cal O_\mu.
\tag{\eqlab{KPol}}$$
As a polarization, $v_k$ thought of as vectors in $T_\mu\Or_\mu$
satisfies
$$\O_{\Or_\mu}(v_h,v_k)=0\tag{\eqlab{vk}}$$
where $\O_{\Or_\mu}$ is the canonical symplectic form on $\Or_\mu$.
The specifics of the choices of $v_k$ will not be important in what follows.
It is worth mentioning that $V_k$ is contained in the vertical polarization
on $T^*N$.

Consider the complex distribution on $N^\#\times\Cal O_\mu$ generated by
$\{\frac{\p\ }{\p p_a},V_k\}$, it is $G-$equivariant thus projects onto
$N^\#\times_G\Cal O_\mu\simeq N^\#/H\times\{\mu\}$. One checks that the 
image is a polarization which we denote by $\P$ \cite{\reflab{Robson}}. 

It is easy to represent 
$\P$ in local coordinates on $N^\#$; $\frac{\p\ }{\p p_a}$ are vector fields
on $N^\#$, and $V_k$ corresponds to $\hat n_*v_k$ as complex vector fields
on $N$, $\hat n_*$ as in (\eqlab{basic}).
This is
so since the assignment $(p,n,\text{Ad}_g\mu)\mapsto(p,ng,\mu)$ defines
the projection $N^\#\times\Cal O_\mu\to N^\#\times_G\Cal O_\mu \simeq
N^\#/H\times\{\mu\}$. Thus vector field generated by $G-$action on $\Cal O_\mu$
translates to vector field generated by $G-$action on $N$.

The Hilbert space structure on $L^2(N)$ is given by integration with
respect to the measure $\d n=\d\sigma\sqrt{\det(g)}\,\d q$, where $\d\sigma$
is a Haar measure on $G$ which we transfer to a measure on the fiber
in the projection $N\to Q$ and $g$ is the metric on $Q$. Using 
the half-form bundle formalism the wavefunctions are
of the form $\psi(n)\sqrt{\d n}$. 
It is clear the the Haar measure will play no role in our consideration
as the polarization and all Hamiltonians in question are $G-$invariant.
To keep the half-form bundle formalism to a minimum, we may identify the
wavefunctions as $\psi(n)\det g^{1/4}$.
We will determine explicitly the 
differential operators corresponding to $f$ so that
$$\psi(n)\det g^{1/4}\mapsto [\Cal Q(f)\psi(n)]\det g^{1/4}.
\tag{\eqlab{simtrans}}$$
In quantizing $f$ that is linear in the momentum variables,
the $\det g^{1/4}$ term will give rise to the covariant divergence,
and for $f=\tfrac{1}{2}||p||^2+V(q)$, it results in the Ricci curvature.
The appearance
of the Ricci curvature is also reported in \cite{\reflab{woodhouse}}.

\subheading{V. Polarization preserving case}

If $f:T^*Q\to \bold R$ is linear in $p$, 
$f=K_a(q)p_a$, then one easily checks, using
(\eqlab{explicitham}), that $\exp t\Ham_{f*}\P=\P$. In fact, we have
\proclaim{Proposition \thmlab{hamfieldprop}}
$$\align\left[\Ham_f,\frac{\p\ }{\p p_a}\right]&=
\frac{\p K_a}{\p q_b}\frac{\p\ }{\p p_b}\,,\tag{\eqlab{braone}}\\
\left[\Ham_f,V_k\right]&\in \text{span}\left\{\frac{\p\ }{\p p_a}\right\},
\tag{\eqlab{bratwo}}
\endalign$$
where the brackets refer to the Lie algebra bracket on vector fields.
\endproclaim
\demo{Proof}
Equation (\eqlab{braone}) is by direct computation. We have
$$\Ham_f=\left(E_b-p_c\frac{\p K_c}{\p q_b}\right)\frac{\p\ }{\p p_b}
+K_b M_{ib}\frac{\p\ }{\p n_i}\tag{\eqlab{hampreserve}}$$
where $E_b$ is independent of $p_a$, and (\eqlab{braone}) results.

To show (\eqlab{bratwo}), we first realize from Proposition \thmlab{function}
that
$\Ham_f=W_1+W_2$ where $W_1$ is the horizontal lift of a vector field on
$T^*Q$, thus $[W_1, V_k]=0$ as $V_k$ is generated by the group action on $N$.
$W_2$ is of the form $F_a\tfrac{\p\ }{\p p_a}$. Since $V_k$ is independent 
of $p$, $[W_2, V_k]=V_k(F_a)\tfrac{\p\ }{\p p_a}$, where $V_k(F_a)$ refers to
applying the vector field as a differential operator to the coefficient
function $F_a$.\qed
\enddemo
The importance of (\eqlab{bratwo}) is that $[\Ham_f, V_k]$ is a 
combination of vectors fields in $\P$ which does not 
involve the $V_h$ vector fields. 
According to 
(7.12) of \cite{\reflab{Sn}}, the quantization of $f$ is then given by:
$$\align
\Cal Q(f)\psi&=-\frac{i\hbar}{\det g^{1/4}}\left[
\Ham_f(\psi(n) \det g^{1/4})+\frac{1}{2}
\sum_{a=1}^{\dim Q}\frac{\p K_a}{\p q_a}(\psi(n) \det g^{1/4})\right]
\tag{\eqlab{preoper}}\\
&=-i\hbar\left[\Ham_f(\psi) + \frac{1}{2}\left(
\frac{1}{\sqrt{\det g}}\Ham_f(\sqrt{\det g})
+\sum_{a=1}^{\dim Q}\frac{\p K_a}{\p q_a}\right)\psi\right]
\endalign$$
The Hamiltonian vector field $\Ham_f$ projects to a 
vector field $V^\#$ on $N$, which is the horizontal lift to the projection
of $\Ham_f$ onto $Q$ where $V=K_b\frac{\p\ }{\p q_b}$. Note that 
$\Ham_f(\sqrt{\det g})= V(\sqrt{\det g})$.
The divergence of the vector field $V$ on $Q$ is defined \cite{\reflab{CWD}}
through the relation
$$\d*V=\text{div}\,V\,\sqrt{\det g}\,\d q
$$
The covariant divergence on $N$ is defined as
the divergence of the horizontal lift $V^\#$. We have
$$\text{div}\,V
=\frac{1}{\sqrt{\det g}}\ V(\sqrt{\det g})+
\sum_{a=1}^{\dim Q}\frac{\p K_a}{\p q_a}\,.$$
Using (\eqlab{hampreserve}), (\eqlab{preoper}) and the fact that
$\psi$ is independent of $p$, we have
\proclaim{Proposition \thmlab{preserve}} 
$\Cal Q(f)\psi=-i\hbar(K_a\D_a+\tfrac{1}{2}\text{div}\, V)
\psi$
where $\D$ is the covariant derivative with respect to the connection $\a$.
\endproclaim
Since $\det g$ is a function of $n$ through $q$, the divergence and the 
covariant divergence are the same.

\subheading{VI. BKS pairing case}

Let $\P$ and $\P'$ be transversal polarizations on $T^*N$, denote their
associated quantum spaces by $\Cal Q$ and $\Cal Q'$. The BKS pairing
gives rise to a map $B:\Cal Q'\to\Cal Q$ such that
$$\langle B(\psi),\phi\rangle = 
\int_{T^*N}\psi \bar\phi (\det\omega)^{1/2} \d\ell$$
where $\d\ell$ is the Liouville form $\d\xi_1\dots\d\xi_n\d n_1\dots\d n_n$
Since the volume form on $N$ is $\d n=\sqrt{\det g}\,\d n_1\dots\d n_n$, 
we have
$$B\psi(n) =\frac{1}{\sqrt{\det g}}\int_{T^*_nN}\psi (\det\omega)^{1/2}
\d\xi_1\dots\d\xi_n\tag{\eqlab{BKS}}$$

Let 
$f=\tfrac{1}{2}g^{ab}p_a p_b+V(q)$, where $g^{ab}$ is the inverse
of the metric $g_{ab}$. From (\eqlab{explicitham}) we
have
$$\align
\Ham_f=&-\left(\frac{\p V}{\p q_a}+\frac{1}{2}
\frac{\p g^{bc}}{\p q_a}p_bp_c\right)
\frac{\p\ }{\p p_a}
+M_{ia}g^{ab}p_b\frac{\p\ }{\p n_i}\\
&+\mu_sM_{ia}M_{jb}
\left(\frac{\p A_{sj}}{\p n_i}-\frac{\p A_{si}}{\p n_j}\right)
g^{bc}p_c\frac{\p\ }{\p p_a}\,,\endalign$$
and note the linear dependence of the coefficients on the $p$ variables.
We denote by $\P_t=\exp t{\Ham_f}_*\P$, here the two
polarizations $\P$ and $\P_t$ do not
intersect transversely. We claim
\proclaim{Proposition \thmlab{polar}} 
Vector fields generated by the group action
are in $\P\cap\P_t$.\endproclaim
\demo{Proof} It suffices to show that 
$$\align \O(\exp t\Ham_{f*}V_k,\frac{\p\ }{\p p_a})&=0\tag{\eqlab{condone}}\\
\text{ and }\qquad
\O(\exp t\Ham_{f*}V_k, V_h)&=0\tag{\eqlab{condtwo}}
\endalign$$ 
with $\O$ as in (\eqlab{localform}).
Let $\bar p(p,n,t)$ and $\bar n(p,n,t)$ denote the flow generated by
$\exp t\Ham_f$ at $(p,n)\in N^\#$ with $\mu$ fixed, 
$(\bar p,\bar n, \mu)=\exp t\Ham_f(p,n,\mu)$.
Then
$\exp t\Ham_{f*}V_k=V_k(\bar p_a)\frac{\p\ }{\p p_a}+V_k(\bar n_i)
\frac{\p\ }{\p n_i}$. So $\O(\exp t\Ham_{f*}V_k,\frac{\p\ }{\p p_a})=
\frac{\p q_a}{\p n_i}V_k(\bar n_i)$. Recall $V_k=\hat n_*v_k$ is 
vertical and $\frac{\p q_a}{\p n_i}$ is the Jacobian of the projection
$\pi:N\to Q$. Then $\pi_*\hat n_*=0$ implies (\eqlab{condone}) holds.

Equation (\eqlab{condtwo}) follows from general principle;
Since $f$ is $G-$invariant (with respect to the non-canonical $G-$action), 
$V_h$ is equivariant with respect to the flow:
$\exp t\Ham_{f*}V_h(p,n)=V_h(\exp t\Ham_f(p,n))$. Since the
flow of a Hamiltonian vector field preserves the symplectic form, we have 
$$\O(\exp t\Ham_{f*}V_k(p,n),V_h(\exp t\Ham_f(p,n)))=
\O(V_k,V_h)=\O_{\Or_\mu}(v_h,v_k)=0,$$ 
where $v_h$ and $v_k$ belongs to a 
polarization on $\Or_\mu$
to begin with (\eqlab{vk}).\qed
\enddemo
This being the case, quantization of $f$ via BKS pairing involves integrating
only over the $p$ variables, \ie, the fiber coordinates of the projection
$\Pi:N^\#\to N$. According to (7.20) of \cite{\reflab{Sn}}, together with
the similarity transform (\eqlab{simtrans}) adjustment and the adjustment 
in the BKS pairing described in (\eqlab{BKS}),
$$\align\Cal Q(f)\psi(n)&=\frac{1}{\det g^{1/2}}\frac{1}{\det g^{1/4}}
i\hbar\left.\frac{\d\ }{\d t}\right|_{t=0}\Psi_t(n)\tag{\eqlab{Quanf}}\\
\text{where}\quad \Psi_t(n)&=(i\hbar)^{-\dim Q/2}\int_{\Pi^{-1}(n)}
\left[\det\omega_{ab}\right]^{1/2}\,\exp(i\hbar^{-1}L)
\,\Psi(p,n,t)\,\text{d}p,\tag{\eqlab{psit}}\\
\Psi(p,n,t)&=\psi(\bar n(p,n,t))\times [\det g(\bar n(p,n,t))]^{1/4},
\tag{\eqlab{Psi}}\\
\omega_{ab}&=\O\left(\frac{\p\ }{\p p_a},
\exp t\Ham_{f*}\frac{\p\ }{\p p_b}\right),\tag{\eqlab{term}}\\
L&=t\left(\tfrac{1}{2}||p||^2+V(q)\right)
-2\int_0^tV(\bar n(p,n,s))\,\d s.\tag{\eqlab{Lagr}}\endalign$$ 

The manipulation  follows closely that of
Sniatycki \cite{\reflab{Sn}}. Making the substitution $x_a=tp_a$,
 we have results analogues to (7.26) and
(7.27) of \cite{\reflab{Sn}}:
\proclaim{Proposition \thmlab{lim}}
$$\align\lim_{t\to 0+}t^{-\dim Q/2}\exp\left(\frac{i}{\hbar}
\frac{||x||^2}{2t}\right)&=
(2\pi\hbar)^{\dim Q/2}e^{\pi i\sgn(g)/4}\sqrt{\det g}\ \delta(x).
\tag{\eqlab{delta}}\\
\frac{\p\ }{\p t}\,t^{-\dim Q/2}\exp\left(\frac{i}{\hbar}
\frac{||x||^2}{2t}\right)&=
\frac{i\hbar}{2}g_{ab}\frac{\p^2\ }{\p x_a\p x_b}\,
t^{-\dim Q/2}\exp\left(\frac{i}{\hbar}\frac{||x||^2}{2t}\right).
\tag{\eqlab{diff}}
\endalign$$
\endproclaim 
\demo{Proof} The first equation
 follows from the method of stationary phase (cf.
\cite{\reflab{GA}}), which for $n-$dimensional space reads
$$\int_{\bold R^n} a(y)e^{ik\phi(y)}\,\d y=
\left(\frac{2\pi}{k}\right)^{n/2}\sum_{y|\d\phi(y)=0}e^{\pi i\sgn H(y)/4}
\frac{e^{ik\phi(y)}a(y)}{\sqrt{|\det H(y)|}}+O(k^{-n/2-1}).$$
The $\tfrac{1}{t}$ factor in (\eqlab{delta})
plays the role of the large parameter $k$. $H$ is the
Hessian of $\phi$ which in our case is $g^{\mu\nu}$, thus $\det H=\det g^{-1}$
and \sgn\ is the signature of the metric. The only stationary 
point in (\eqlab{delta}) is $x=0$, thus the right hand side of
(\eqlab{delta}) has a ($\dim Q-$dimensional) delta function at $x=0$.

The second equation is a straight forward computation. 
We need the fact that 
$g_{a\mu}g^{\mu b}=\delta_{ab}$, and
$\sum_a\sum_b g_{ab}g^{ab}=\sum_a\delta_{aa}=\dim Q$ in the course of the
computation.\qed
\enddemo
One checks that $\omega_{ab}$ in (\eqlab{omega}) is
$$\frac{\p q_a}{\p  n_i}\frac{\p \bar n_i}{\p p_b}=
tg^{ab}+\text{higher order terms in }t$$
using (\eqlab{nbar}) below.
Then $[\det\omega_{ab}]^{1/2}\d p\sim t^{-\dim Q/2}\d x$, 
providing us with the needed
factor to apply the results of Proposition \thmlab{lim}.
Thus in determining $\left.\tfrac{\d\ }{\d t}\right|_{t=0}\Psi_t(n)$, 
we need only to
consider terms involving $t$, $tp_a$ and $t^2p_ap_b$ while
ignoring terms of the form $t^2p_a$ and all 
higher order
terms. Using (\eqlab{explicitham}),
the expansion of $\bar n$, expressed in the $t$ and
$x_a$ variables, up to the relevant terms are
$$\align
\tilde n_i(n,x)=&n_i+M_{ia}g^{\mu a}x_\mu
+\frac{1}{2}\left[
M_{jb}\left(\frac{\p M_{ia}}{\p n_j}g^{\nu b}g^{\mu a}\right.\right.\\
&\left.\left.+M_{ia}\frac{\p g^{\mu a}}{\p q_b}g^{\nu b}\right)
-\frac{1}{2}M_{ia}\frac{\p g^{\mu\nu}}{\p q_b}g^{ba}\right]x_\mu x_\nu
\tag{\eqlab{nbar}}\endalign$$
which is independent of $t$.
And $\Psi_t(n)$ in (\eqlab{psit}) is reduced to
$$\align
\Psi_t(n)&=(i\hbar)^{-\dim Q/2}\int_{\Pi^{-1}(n)}
t^{-\dim Q/2}\exp\left(\frac{i}{\hbar}\frac{||x||^2}{2t}\right)
\Phi(n,x,t)\,\d x,\\
\text{where }&\quad
\Phi(n,x,t)=\exp(-i\hbar^{-1}tV(q))\times\psi(\tilde n(n,x))
\det g(\tilde n(n,x))^{1/4}.\tag{\eqlab{Phinew}}\endalign$$
By applying Proposition \thmlab{lim}, integration by parts yields
$$\Cal Q(f)\psi(n)=\frac{(-2\pi i)^{\dim Q/2}e^{\pi i\sgn/4}}{\det g^{1/4}}
\left[-\frac{\hbar^2}{2}g_{ab}
\left.\frac{\p^2\Phi}{\p x_a\p x_b}\right|_{x,t=0}
+i\hbar\left.\frac{\d\ }{\d t}\right|_{x=t=0}\Phi\right]\tag{\eqlab{Qexpl}}$$
The $\sqrt{\det H}$ term that appears in the method of stationary phase
formula results in a $g^{1/2}$ factor (\eqlab{delta}) that cancels with 
the $\det g^{1/2}$
on the right hand side of (\eqlab{Quanf}).

Since $n$ is fixed, we can choose a normal coordinate system 
\cite{\reflab{CWD}} around $q=\pi(n)$
so that $\tfrac{\p g^{\mu\nu}}{\p q_a}=0$ for all $\mu,\nu$ and $q_a$ when
evaluated at $n$, (\ie, at $x=t=0$). A direct computation shows
$$\align
\left.\frac{\d\ }{\d t}\right|_{x=t=0}\Phi(n,x,t)
&=-i\hbar^{-1}V(q)\psi(n)\det g^{1/4}\tag{\eqlab{dPhidt}}\\
g_{ab}\left.\frac{\p^2\ \ }{\p x_a\p x_b}\right|_{x=t=0}
\Phi(n,x,t)&=\det g^{1/4}\left[g^{\mu\nu}M_{j\nu}
\frac{\p M_{i\mu}}{\p n_j}\frac{\p\psi}{\p n_i}\right.\tag{\eqlab{pgpxx}}\\
&\ +\left.
g^{\mu\nu}M_{i\mu}M_{j\nu}\frac{\p^2\psi}{\p n_i\p n_j}
+\frac{1}{4}g^{\mu\nu}g^{ab}\frac{\p^2g_{\mu\nu}}{\p q_a\p q_b}\right].
\endalign$$
Here we have made repeated use of the identity \cite{\reflab{CWD} p.302}, 
$$\frac{1}{\det g}\frac{\p\det g }{\p q_b}=
\frac{1}{\det g}\frac{\p\det g}{\p g_{\mu\nu}}\frac{\p g_{\mu\nu}}{\p q_b}
=g^{\mu\nu}\frac{\p g_{\mu\nu}}{\p q_b}.$$

In normal coordinates, the covariant Laplace operator reduces to
$$\align\Delta_\alpha\psi&=\frac{1}{\det g^{1/2}}
\D_\mu(\det g^{1/2}g^{\mu\nu}\,\D_\nu\psi)\\
&=g^{\mu\nu}M_{i\mu}M_{j\nu}\frac{\p^2\psi}{\p n_i\p n_j}
+ g^{\mu\nu}M_{j\nu}
\frac{\p M_{i\mu}}{\p n_j},\tag{\eqlab{Laplace}}\endalign$$
and the Ricci curvature becomes
$$\align R&=g^{ik}(\p_k\Gamma_{ji}^j-\p_j\Gamma_{ki}^j
+\Gamma_{km}^j\Gamma_{ji}^m-\Gamma_{jm}^j\Gamma_{ki}^m)\\
&=(g^{ik}g^{\mu\nu}-g^{i\nu}g^{k\mu})\p_i\p_kg_{\mu\nu}\\
&=\tfrac{3}{2}g^{ik}g^{\mu\nu}\p_i\p_kg_{\mu\nu}\tag{\eqlab{Ricci}}\endalign$$
Here $\Gamma_{ij}^k$ are the Christoffel symbols, and the identity
$$\Gamma_{ij,k}^m+\Gamma_{jk,i}^m+\Gamma_{ki,j}^m=0$$
is used to show 
$g^{i\nu}g^{k\mu}\p_i\p_kg_{\mu\nu}
=-\tfrac{1}{2}g^{ik}g^{\mu\nu}\p_i\p_kg_{\mu\nu}$.
We must caution the readers that these expressions only hold in
normal coordinates. However, by combining (\eqlab{Qexpl}--\eqlab{Ricci}) 
we can express our
final result in an coordinate invariant form:
\proclaim{Proposition \thmlab{main}}
Quantization of $f=\tfrac{1}{2}||p||^2+V(q)$ gives
$$\Cal Q(f)\psi(n)=
(-2\pi i)^{\dim Q/2}e^{\pi i\sgn(g)/4}
\left(-\frac{\hbar^2}{2}
\left[\Delta_\a+\frac{1}{6}R\right]+V(q)\right)\psi(n).$$
\endproclaim

We conclude with a final remark. 
The 
Yang-Mills field is defined \cite{\reflab{WuYang}} as the curvature $\D\a$
of the 
Yang-Mills potential $\a$, whereas the contribution of this connection 
in the local expression of the symplectic form is $\d\a$. They are related by
$$\D\a(v,w)=\d\a(\hor v,\hor w)$$
where $\hor$ denotes the horizontal projection. Since the vector fields
of concern are all horizontal, the effect of $\d\a$ is equivalent to the
curvature.

\subheading{Acknowledgement} We wish to thank N\.P\. Landsman for helpful
comments on the preliminary version of this work.
\vfil\pagebreak
\centerline{\bf References}
\vskip.4cm

\widestnumber\key{WW}

\ref \no\reflab{MW}
\by J\. Marsden and A\. Weinstein
\pages 121(1974)
\vol 5
\jour Rep\. Math\. Phys\.
\endref

\ref \no\reflab{Weinstein}
\by  A\. Weinstein
\pages 417(1978)
\vol 2
\jour Lett\. Math\. Phys\.
\endref

\ref \no\reflab{Sn}
\by J\. \'Sniatycki
\book Geometric quantization and quantum mechanics
\publ  (Springer-Verlag, New York, 1980)
\endref

\ref \no\reflab{woodhouse}
\by N\.M\.J\. Woodhouse
\book Geometric Quantization
\publ (Clarendon Press, Oxford, second edition, 1992)
\endref

\ref \no\reflab{Ki}
\by A\.A\. Kirillov
\book Elements of the theory of representation
\publ  (Springer-Verlag, Berlin, 1976)
\endref

\ref \no\reflab{Ko}
\by B\. Kostant
\paper Quantization and unitary representation. In: Modern analysis
and applications
\pages 87-207
\yr 1970
\vol 170
\jour Lecture Notes in Math\.
\endref

\ref \no\reflab{multi}
\by V\.W\. Guillemin and S\. Sternberg
\pages 515(1982)
\vol 67
\jour Invent\. Math\.
\endref

\ref \no\reflab{Robson}
\by  M\.A\. Robson
\pages 207(1996)
\vol 19
\jour J\. Geom\. Phys\.
\endref

\ref \no\reflab{BKS}
\by R\.J\. Blattner
\pages 87(1973)
\vol 26
\jour Proc\. Symp\. Pure Math\.
\endref

\ref \no\reflab{Landsman}
\by  N\.P\. Landsman
\pages 93(1993)
\vol 12
\jour J\. Geom\. Phys\.
\endref

\ref \no\reflab{Kummer}
\by M\. Kummer
\pages 281(1981)
\vol 30
\jour Indiana Univ\. Math\. J\.
\endref

\ref \no\reflab{CWD}
\by Y\. Choquet-Bruhat, C\. de Witt-Morette and M\. Dillard-Bleick
\book Analysis, manifolds and physics
\publ (North-Holland, Amsterdam, 1977)
\endref

\ref \no\reflab{GA}
\by V\.W\. Guillemin and S\. Sternberg
\book Geometric asymptotics
\publ  (Amer\.Math\.Soc\., Providence, 1977)
\endref

\ref \no\reflab{WuYang}
\by  T\.T\. Wu and C\.N\. Yang
\pages 3845(1975)
\vol 12
\jour Phys\. Rev\. D
\endref

\end